\documentclass[aps,pra,amsmath,amssymb,superscriptaddress,floatfix,showpacs,twocolumn]{revtex4-2}

\usepackage{graphicx}
\usepackage{mathdots}
\usepackage{epstopdf} 
\usepackage[usenames,dvipsnames]{color}
\usepackage{bm}
\usepackage{inputenc}
\usepackage{xcolor}
\usepackage{tikz}
\usepackage{physics}
\usepackage{booktabs}

\usepackage[colorlinks,bookmarks=false,citecolor=blue,linkcolor=red,urlcolor=blue]{hyperref}
\usepackage[colorlinks,bookmarks=false,citecolor=blue,linkcolor=red,urlcolor=blue]{hyperref}
\usepackage{xcolor}
\usepackage{mathdots}
\usepackage{float}
\usepackage{needspace}
\usepackage{soul}
\usepackage[usenames,dvipsnames]{color}

\begin{document}
\preprint{APS/123-QED}
\title{Nonlinear dynamics in an artificial feedback spin maser}
\author{Mingjun Feng}
\author{Lan Wu}
\affiliation{National Time Service Center, Chinese Academy of Sciences, Xi'an, 710600, China.}
\affiliation{University of Chinese Academy of Sciences, Beijing, 100049, China}
\author{Guobin Liu}
\email{liuguobin@ntsc.ac.cn}
\affiliation{National Time Service Center, Chinese Academy of Sciences, Xi'an, 710600, China.}
\affiliation{University of Chinese Academy of Sciences, Beijing, 100049, China}
\affiliation{Key Laboratory of Time Reference and Applications, Chinese Academy of Sciences, Xi’an 710600, China}

\date{\today}
\begin{abstract}
Spin masers with optical detection and artificial feedback are widely used in fundamental and practical applications. However, a full picture of the maser dynamics is still absent. By solving the feedback driven Bloch equations, we simulated the dynamics of an ideal spin maser in a broad parameter space. Rich nonlinear dynamics including high order harmonics generation, nonperiodic spin oscillations and frequency comb were revealed when the artificial feedback interaction exceeds the normal spin-field interaction. We also propose a pulse feedback spin maser protocol, which constructs an ultralow field magnetic frequency comb and could be useful in precision atomic magnetometers in searching for spin-dependent exotic interactions.
\end{abstract}
\maketitle

The idea of spin maser was conceived at the time of laser's invention \cite{Dehmelt1957,Bloom1962} and the first $^3$He maser was soon realized \cite{Robinson1964}. A thorough theoretical analysis of maser dynamics was given within the framework of Bloch equations in nuclear magnetic resonance (NMR) and potential applications of spin masers in magnetometer and NMR gyroscope were also forecasted \cite{Richards1988}. Since then, spin maser had been realized with various types, e.g. single frequency spin maser, dual-frequency spin maser and modulated spin maser \cite{Chupp1994,Stoner1996,Bear1998,Yoshimi2002,Yoshimi2012,Sato2018,Bevington2020,Jiang2021,Wang2023PRApplied} and used for fundamental physics, such as tests of CPT symmetry, searches for permanent EDM in neutral atoms $^3$He and $^{129}$Xe and searches for dark matters \cite{Bear2000,Rosenberry2001,Jiang2021,Afach2021}. Low sensitivity and inexpensive alkali-metal spin maser can be used for non-destructive tests of defects in metal materials \cite{Bevington2020,Bevington2019}.

Pick-up coil detection is widely used in conventional spin masers. Due to Faraday's law of electromagnetic induction, the operational frequency of spin maser has to be maintained at relatively high values, i.e. several kilohertz \cite{Chupp1994,Stoner1996,Bear1998}. Pick-up coil serves also as an excitation coil for continuous spin oscillations, thus high quality magnetic coils with fine-tunable parameters are the key ingredients in realizing effective spin masers. Optical detection can help to share the role of pick-up coils and was first realized with a probe laser in an artificial feedback spin maser \cite{Yoshimi2002}. The use of optical detection not only improves the detection efficiency, but also allows for a lower maser frequency down to Hertz level in contrast to kilohertz operation of previous spin masers. This is very important in applications requiring an absolute energy or frequency resolution such as searches for EDM and precision measurement of inertial rotation.

Radiation damping time of pick-up coils plays an important role in revealing the masing mechanism of spin masers \cite{Richards1988}. While the terminology of radiation damping time originates from NMR spectroscopy and does not conflict with that of artificial feedback field in explaining the optically detected spin maser \cite{Yoshimi2002}, it may not reveal its whole details. A complete understanding on the feedback effects in maser dynamics is still lacking albeti recent simulation works \cite{Wang2017,Li2023}. On the other hand, systematic theoretical investigation on the spin maser dynamics is urgently needed as the ability to control feedback parameters experimentally advances \cite{Jiang2021,Li2023}.

Following the methodology of artificial feedback spin maser, here we try to obtain a complete map of the maser dynamics. We simulate the spin maser dynamics theoretically using the time-delay feedback \cite{Balachandran2009} and present a full map of the spin maser dynamics depending on the strength and phase delay of the feedback field. New effects such as the high order harmonics generation similar to the multiphoton process in atom-photon interactions and nonperiodic spin oscillations similar to the chaos, emerge as the artificial feedback interaction exceeds the spin-field interaction. Furthermore, we propose a new kind of spin maser by introducing a pulse time sequence in the feedback protocol. 

\textit{Model--}Consider a single species and single isotope (either alkali metal or noble gas, e.g. $^{87}$Rb or $^{129}$Xe) with spin magnetization $\mathbf{M}=(M_x, M_y, M_z)$ moving in a field $\mathbf{B}$. The equation of motion of the spin system can be described by Bloch equations
\begin{equation}\label{eq:DynamicsEquations}
\begin{aligned}
\dot M_x&=\gamma(M_yB_z-M_zB_y)-\frac{M_x}{T_2}\\
\dot M_y&=\gamma(M_zB_x-M_xB_z)-\frac{M_y}{T_2}\\
\dot M_z&=\gamma(M_xB_y-M_yB_x)+\frac{M_0-M_z}{T_1},
\end{aligned}
\end{equation}
where the total field is $\mathbf{B}=(B_x,0,B_0)$, including a static longitudinal field $B_0$ in the $z$ direction and a transverse field $B_x$ in the $x$ direction. $\gamma$ is the gyromagnetic ratio, $T_1$ and $T_2$ are longitudinal and transverse relaxation times, $M_0$ is the equilibrium magnetization created by optical pumping and relaxation when feedback is absent.
\begin{figure}[h]
	\centering
	\includegraphics[scale=1]{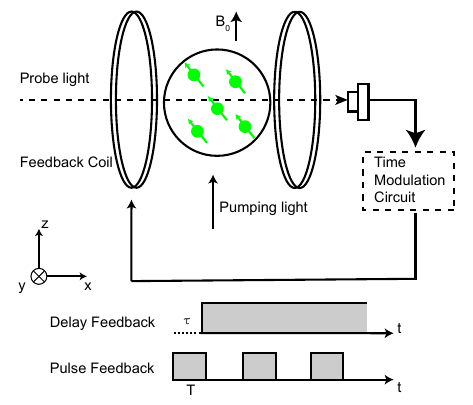}
	\caption{Setup of a spin maser with optical detection and artificial feedback. Atomic spins are polarized by a pump light and precess around the bias magnetic field $B_0$ along the $z$ axis. The transverse magnetization $M_x(t)$ is detected by a probe light and then processed by a homemade electronic circuit. The processed signal is connected to a feedback coil, providing the feedback magnetic field $B_x$. For delay feedback, $B_x$ is given by Eq.~\ref{eq:delayfeedback}. For pulse feedback, $B_x$ is given by Eq.~\ref{eq:pulsefeedback}.}
	\label{fig:set up}
\end{figure}

The optically detected spin maser with artificial feedback is illustrated in Fig. \ref{fig:set up}. Here we define two kinds of spin masers, delay feedback and pulse feedback spin masers. For the delay feedback spin maser, the transverse feedback field $B_x$ is
\begin{equation}
B_x= {k  M_x(t- \tau)}, 
\label{eq:delayfeedback}
\end{equation}
with $k$ the feedback coefficient and $\tau$ the time delay.
To give $\tau$ a specific meaning, we set $\tau=\psi/\omega_0$, where $\omega_0=\gamma B_0$ is the Larmor precession frequency and $\psi$ is the phase delay corresponding to $\tau$.
For the pulse feedback spin maser, the $B_x$ field is
\begin{equation}
B_x =
\begin{cases}
kM_y(t), \quad 2nT \leqslant t \leqslant (2n+1)T\\
0, \quad (2n+1)T<t<(2n+2)T,
\end{cases}
\label{eq:pulsefeedback}
\end{equation}
where $k$ is the feedback coefficient, $n$ is a positive integer, $T$ is the duration of the pulse and also the interval time between two pulses.

In this paper, we take an ideal $^{129}$Xe spin maser as the model. In practice, $^{129}$Xe spins can be hyperpolarized to a high degree via spin-exchange collisions with media spins (Rb electron spins for example), manipulated by magnetic fields and detected by a probe light \cite{Chupp1994,Yoshimi2002,Jiang2021,Li2023}. Here in the ideal Xe spin maser model, we ignore the Rb-Xe spin coupling dynamics and consider the Xe spin dynamics only. The parameters used in simulations are, gyromagnetic ratio $\gamma = -2\pi \times 11.78 \ \text{rad} \cdot \text{MHz/T} $ \cite{Ratcliffe1998}, bias static magnetic field $B_0 = 3 \times 10^{-6}$ T, thus the corresponding Larmor frequency is $\nu_0 = |\gamma B_0|/(2\pi) = 35.34$ Hz. The longitudinal and transverse spin relaxation time is $T_1$ and $T_2$ s, respectively, both of which are set to 10 seconds and the initial $^{129}$Xe spin magnetization is $M_0 = 4$ A/m. For both delay and pulse feedback spin masers, we define $k' = k M_0/B_0$, which is easy to evaluate the strength of the feedback field relative to that of the static magnetic field.

For delay feedback in Eq.\ref{eq:delayfeedback}, the solutions of Eq. (\ref{eq:DynamicsEquations}) are determined mainly by two parameters, $k'$ and $\psi$. Given the strength of the feedback field relative to that of the static field, the full regime of spin maser dynamics can be generally divided into two regimes: the weak feedback with $|k'| \le 1$ and strong feedback regimes with $|k'| \ge 1$. Considering the different phase delay of the feedback field, i.e. $\psi \in [0,\pi]$ for forwarded delay or $\psi \in [\pi, 2\pi]$ for reversed delay, the maser dynamics can be further classified as four regimes. We will discuss the details in the following.
\begin{figure}[h]
	\centering
	\includegraphics[scale=0.8]{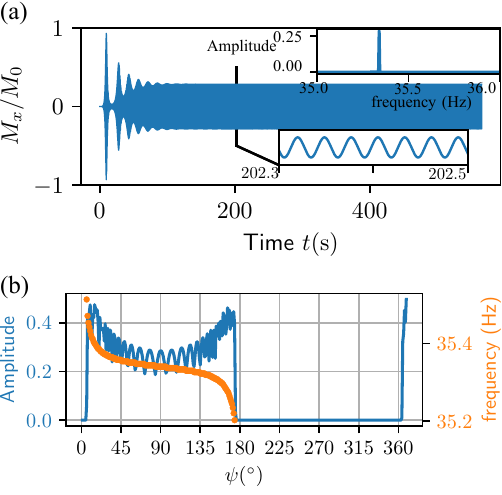}
	\caption{(a) Transverse magnetization $M_x(t)$ oscillations with time at $k'=-0.01$ and $\psi=90^\circ$. (b) Frequency $\nu$ and amplitude response of spin maser as a function of the phase $\psi=\omega_0 \tau$ at $k'=-0.01$.}
	\label{fig:vary_psi}
\end{figure}

\begin{figure*}[ht]
	\centering
	\includegraphics[scale=1]{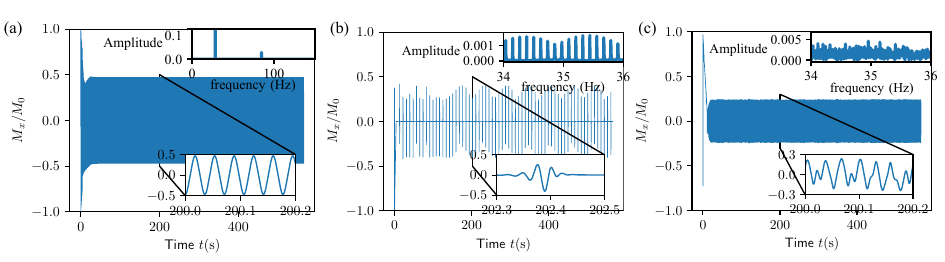}
	\caption{When the artificial feedback interaction is much stronger than the spin-field interaction, various nonlinear spin dynamics occur:(a) $k'=-2, \psi=225^\circ$, high order harmonics generation, (b) $k'=-2, \psi=90^\circ$, frequency comb and (c) $k'=-10, \psi=180^\circ$, nonperiodic spin oscillations.}
	\label{fig3}
\end{figure*}

\textit{Weak feedback regime--}First, we try to reproduce the normal spin maser dynamics with a new method of numerical simulation. The normal spin maser dynamics has been simulated previously \cite{Yoshimi2002,Wang2017,Li2023}, where the frequency shift was shown to be linear with the phase delay. As will be shown later, when the feedback strength exceeds critical value, this will be nonlinear with the phase delay or time delay. In this case, previous numerical solving with the Runge-Kutta method was found to be inefficient doing the theoretical simulations \cite{Li2023}. Therefore we solve the equations by using the JiTCDDE (just-in-time compilation for delay differential equations) module \cite{Ansmann2018}, which demonstrates higher efficiency and better reproducibility in dealing with the strong nonlinearity in strong feedback driven Bloch equations.

We set the maximum time step to $\Delta t = 0.00028$ s, and use an initial value $\mathbf{M}(t) = (10^{-4}M_0,0,M_0)$ for $t\in [-\tau, 0]$, which means that the collective spins remain primarily in the equilibrium state but with a tiny tilt away from the $z$ axis to the $x$ axis.
By solving Eqs.~(\ref{eq:DynamicsEquations}) with input parameters $k'=-0.01$ and $\psi=90^\circ$, a typical spin maser oscillation is shown in Fig. \ref{fig:vary_psi}(a). The $M_x (t)$ reaches a steady sinusoidal oscillation at around 100 seconds, with a frequency of 35.34 Hz.
To obtain the frequency and amplitude response of maser signal with respect to the phase delay $\psi$, we fix $k'=-0.01$ and then vary $\psi$ from 0-370$^\circ$. The simulation results are shown in Fig. \ref{fig:vary_psi}(b). 

It is clearly shown that for $\psi \in [0,\pi]$, the results agree well with our previous results with the Runge-Kutta method \cite{Li2023}. The specific phase for masing effect to occur is
\begin{equation}
\sin \psi > \frac{2}{\gamma  k  M_0  T_2} = \frac{2}{\gamma k' B_0 T_2},
\end{equation}
which is consistent with the analytical results of the rotational wave approximation \cite{Wang2017}.
Within the masing phase range, the maser frequency shifts as
\begin{equation}
\delta = \nu-\nu_0 \approx \frac{\tan (\pi/2 - \psi)}{T_2}.
\end{equation}
On the other hand, for $\psi \in [\pi,2\pi]$ and weak feedback strength $k' \le$1, there is no maser effect, which agrees with experiments \cite{Yoshimi2002,Jiang2021,Li2023}. 

\textit{Strong feedback regime--}When the strength of feedback field exceeds that of the main field, i.e. $k' \ge$1, nonlinear effects begin to dominate the maser dynamics. In this regime, there are three distinct behaviors of spin dynamics. For $\psi \in [0, \pi]$, the spin maser becomes like a chirp laser, with pulsed spin oscillations in time domain and comb-like spectra in frequency domain, as shown in Fig.~\ref{fig3}(b). For $\psi \in [\pi, 2\pi]$, there are two kinds of nonlinear dynamics: 1) high order harmonics generation in Fig.~\ref{fig3}(a), and 2) nonperiodic spin oscillations in Fig.~\ref{fig3}(c). 

In Fig.~\ref{fig3}(a), two peaks appear at $\nu$=28 Hz and 3$\nu$=84 Hz in the frequency spectra. The fundamental frequency $\nu$ is obviously different from the Larmor frequency determined by the intrinsic spin-field interaction. Even if the frequency $\nu$ of the transverse excitation field is far detuned from the Zeeman splitting, it can still cause resonance. Furthermore, oscillating magnetic fields with frequencies of $3\nu$ can also cause resonances, corresponding to multiphoton absorption or stimulated radiation processes. In this case, the system is similar to a magnetic resonance system in a strong oscillating magnetic field \cite{Shirley1965, CohenTannoudji1998}. Therefore, feedback spin masers provide a new way of stimulating nonlinear phenomena in atom-photon interaction process.

In Fig.~\ref{fig3}(b), the comb-like spectrum is broadened compared to the original spin maser, providing a new way of creating ultralow field magnetic frequency comb in gaseous spin ensembles other than the magnon frequency comb in solid state systems such as microstructured waveguide and artificial metamaterials \cite{Hula2022,freqcomb1arxiv,freqcomb2arxiv}. Compared to solid state spin systems, gaseous spin systems exhibit less couplings to the external electromagnetic environments, rendering higher sensitivity and stability in precision measurement physics such as atomic magnetometers and gyroscopes. 

In Fig.~\ref{fig3}(c), at even stronger feedback strength, nonperiodic spin oscillations emerge. This is as a natural result due to the nonlinearity in the Bloch equations once including a feedback field term. Actually, chaotic solutions has been numerically simulated by Abergel \cite{Abergel2002} and spatial-temporal chaos had been experimentally observed in an inhomogeneous solution NMR system \cite{Lin2000}, both systems are described by similar feedback driven Bloch equations .  

\begin{figure}[h]
	\centering
	\includegraphics[scale=1]{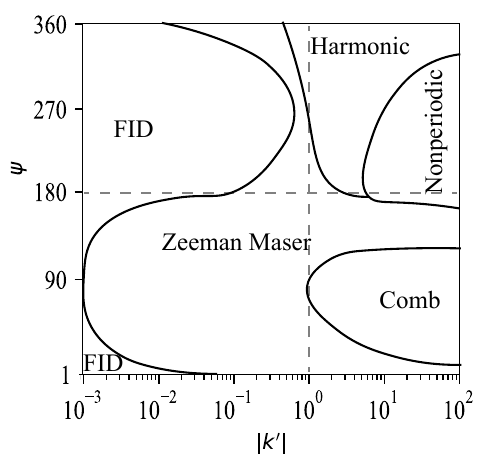}
	\caption{Four regimes of spin maser dynamics: i) FID-free induction decay at weak feedback and reversed phase delay (upper left); ii) Zeeman maser at weak feedback and forwarded phase delay (lower left); iii) High order harmonics and nonperiodic dynamics at strong feedback and reversed phase delay (upper right); iv) Frequency comb at strong feedback and forwarded phase delay (lower right).}
	\label{fig:zone}
\end{figure}

To summarize, Fig.~\ref{fig:zone} presents a full picture of spin maser dynamics in a broad parameter space, i.e. $k' \in -[0.001, 100]$ and $\psi \in [0,2\pi]$. In general, the maser dynamics can be divided into four zones: i) Upper left, FID-free induction decay, where no maser effect occurs due to weak feedback and reversed phase delay; ii) Lower left, Zeeman maser, where normal maser effect occurs due to weak feedback and forwarded phase delay; iii) Upper right, high order harmonics and nonperiodic dynamics, where nonlinear dynamics occurs due to strong feedback and reversed phase delay; iv) Lower right, frequency comb or chirp maser, where spectral broadening and splitting occur due to strong feedback and forwarded phase delay. 

\begin{figure}[h]
	\centering
	\includegraphics[scale=1]{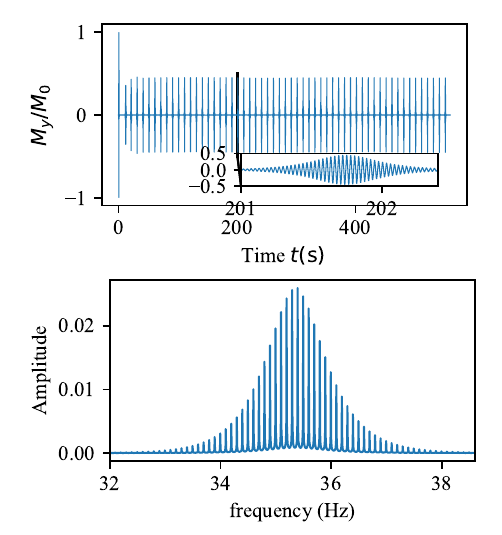}
	\caption{(a) Typical time oscillations of the pulse feedback spin maser with $k'=-0.1$ and $T=5$ s.
	(b) Frequency comb spectrum of the pulse feedback spin maser. The frequency gap between two neighbor comb lines is $\nu_g=1/(2T)=0.1$ Hz.}
	\label{fig:frequency_comb}
\end{figure}

\textit{Pulse feedback spin maser or magnetic frequency comb--}Inspired by the idea of optical frequency comb, we design a pulse feedback spin maser. The protocol is depicted by Fig.~\ref{fig:set up} and Eq.\ref{eq:pulsefeedback}. In this case, equations (\ref{eq:DynamicsEquations}) has no time delay term and depends only on the feedback strength $k$ and pulse duration $T$. We set $k'=0.1$ and $T=5$ s, and the solution $M_y(t)$ is given in Fig. \ref{fig:frequency_comb}. In Fig. \ref{fig:frequency_comb}(a), the spin maser emits short pulses in the time domain. The spectrum represents a typical frequency comb with Larmor frequency $\nu_0 = 35.34$ Hz as the carrier frequency and $\nu_g =  1/(2T) = 0.1$ Hz as the repetition frequency. Different from the optical frequency comb requiring a high quality optical cavity, the pulse spin maser or magnetic frequency comb can be realized in a cavity-free way, simply by switching on and off the feedback circuit. The carrier frequency and repetition frequency of the magnetic frequency comb can be easily tuned by changing the bias magnetic field $B_0$ and the pulse duration $T$. Due to the intrinsically long coherence time of noble gas spins, the magnetic frequency comb can reach high sensitivity and can be useful in searching for spin-dependent exotic interactions such as the axion-like particles \cite{Afach2021}. Compared to the Floquet maser, an external modulation field seems unnecessary in this pulse spin maser protocol \cite{Jiang2021}.

\textit{Conclusion--}In particular, we would like to mention that, the time crystals phenomena in spin maser systems were recently independently reported by Tang et al \cite{Tang2024} and by us \cite{Wang2024}, in pure Rb and Rb-Xe hybrid spin maser systems, respectively. However, we did not find any time crystal effects in the present model of an ideal Xe spin maser system. We have reasons to suspect that the key point of time crystals lies in the different timescales in this system, i.e. the competition between the Larmor periods and spin relaxation times of different spin species. A full theoretical investigation on the Rb-Xe hybrid spin maser dynamics is under way.
 
In conclusion, by simulating time delay spin dynamics of feedback driven Bloch equations in an optically detected spin maser system, we reveal rich nonlinear dynamics of an ideal Xe spin maser. When the feedback field is smaller than the bias magnetic field, the well-known FID and Zeeman maser effects dominate the maser dynamics. However, when the artificial feedback interaction exceeds the spin-field interaction, nonlinear dynamics including high-order harmonics generation, nonperiodic spin oscillations and frequency comb may appear. The direction of phase delay plays an important role in determining the specific nonlinear dynamics of spin masers. We also propose a magnetic frequency comb based on a pulse feedback spin maser protocol. The pulse spin maser can work in an ultralow frequency range as a cavity-free magnetic frequency comb, which might be potentially useful in searching for spin-dependent exotic interactions.

G. L. appreciate the finiancial support by Chinese Academy of Sciences under Grant No. E209YC1101.

%

\end{document}